\documentclass{iaus}
\usepackage{graphicx}

\title[Lensing Statistics and 
$\Omega_{\Lambda}$]{Quasar Lensing Statistics and 
$\Omega_{\Lambda}$:\\ What Went Wrong?}

\author[Dan Maoz]
{Dan Maoz}

\affiliation{School of Physics and Astronomy, Tel-Aviv University, 
Tel-Aviv 69978, Israel; maoz@wise.tau.ac.il}

\pubyear{2004}
\volume{225}
\pagerange{1--8}
\date{?? and in revised form ??}
\jname{Impact of Gravitational Lensing on Cosmology}
\editors{Mellier, Y. \& Meylan,G. eds.}
\begin{document}

\maketitle

\begin{abstract}
In the pre-WMAP, pre-Supernova-Ia-Hubble-diagram era, 
quasar lensing statistics stubbornly
indicated low values of $\Omega_{\Lambda}$. In contrast, a number
of recent lensing statistics studies either
find the data support the standard $\Lambda$CDM picture, or
simply take the standard cosmological parameters as a given. Have the data
or the analyses changed or improved, and how?
 I review several of the ``historical'' and the more 
recent studies, and show that there is no particular 
measurement, assumption, or model parameter in the old studies that
was grossly wrong.
Instead, at least several
effects, operating together, are likely required in order 
to achieve agreement between the observations and 
the currently standard cosmology. Most likely among these effects 
 are: a somewhat lower lensing 
cross section for elliptical galaxies than assumed in the past; some
loss of lensed quasars in optical samples due to extinction by the
lenses; 
and a somewhat lower-than-standard value of $\Omega_{\Lambda}\sim 0.6$.
The agreement between recent model calculations and the results of radio lens
surveys may be fortuitous, and due to a cancellation between the
errors in the input parameters for the lens population and the
cosmology, on the one hand, and for the source population, on the other hand.  

\end{abstract}

\section{Introduction}
The fraction of quasars that are strongly lensed (i.e., split into multiple 
images) by intervening galaxies is a probe of, among other things, the volume
of space between us and the quasars. Since the size of this volume depends 
on cosmological parameters, and in particlular on the cosmological constant,
$\Omega_{\Lambda}$, measurement of the lensed fraction can constrain cosmology.
Specifically, a large value of $\Omega_{\Lambda}$ gives a large
volume, 
out to a given redshift, and 
hence leads to a large lensing probability, assuming 
a constant comoving density of lenses (i.e., galaxies). This idea was first 
outlined in a series by papers of Ed Turner and collaborators (Turner, 
Ostriker, \& Gott 1984; Turner 1990; Fukugita \& Turner 1991; see
Kochanek et al. 2004 for a recent review).
Figure 1 shows the increase in
volume out to a source at $z=2$, as a function of $\Omega_{\Lambda}$,
for flat geometries. 
A useful order of magnitude estimate for the lensing optical depth $\tau$
for sources at distance $D$, lensed by an intervening population with density
$n$, and with each lens having a strong-lensing cross section $\sigma$ is
\begin{equation}
\label{nsigd}
\tau\sim n \sigma D.
\end{equation}
Taking, for $n$, the comoving density of $L_*$ ellipticals, 
$0.5\times 10^{-2}$~Mpc$^{-3}$, for the cross section for multiple lensing,
$\pi R_E^2$, where $R_E\sim 4$~kpc is a typical Einstein radius for an $L_*$
elliptical at $z\sim 0.5$, and a proper-motion distance to the source of
about 4~Gpc ($z_{source}\sim 2$), Eq. \ref{nsigd} gives a lensing otical depth of order $10^{-3}$.
To obtain the lensing probability, the optical depth must be corrected by
the magnification bias, $B$, i.e., the over-representation of lensed objects
in a flux limited sample of sources having a steeply rising number-magnitude 
relation, due to the magnification that lensing entails. For bright quasar 
samples, $B\sim 10$, while for radio samples $B$ is typically a few.
Thus one expects of order 1\% of bright ($\lesssim 18$~mag) quasars, 
and a fraction of a few $\times 10^{-3}$ of radio samples, to be strongly 
lensed. Detailed calculations predict similar numbers.
\begin{figure}
\includegraphics{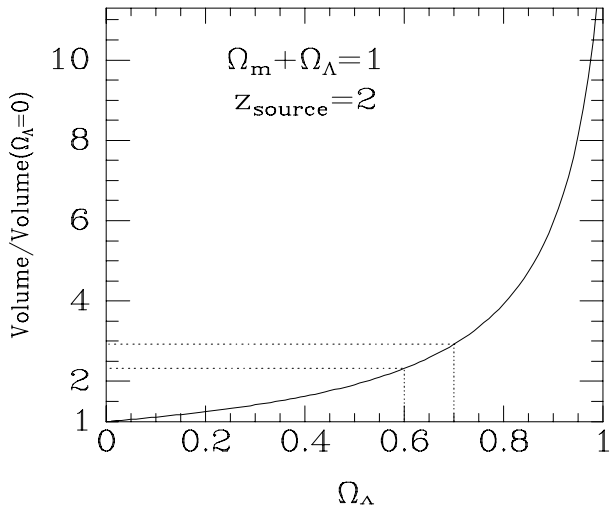}
  \caption{The volume enclosed within the radius out to a source at
    $z=2$,
as a function of $\Omega_{\Lambda}$, relative to this volume for
$\Omega_{\Lambda}=0$, for flat geometries. 
}
\end{figure}

\section{A brief history of lensing surveys and analyses}
The first large optical survey for lensed quasars that was sensitive
over most of the $0''-3''$ range over which galaxy lensing occurs was
the HST Snapshot Survey for lensed quasars (Maoz et al. 1993), which
found that $4/502\approx 1\%$ out of a sample of luminous $z>1$
quasars are lensed. Maoz \& Rix (1993) modeled the results of this
survey using a ``hybrid'' model for galaxies, consisting of a
deVaucouleurs stellar-mass profile, combined with a cored isothermal
sphere distribution representing the dark halo. They found that the 
observed low frequency of lensing was consistent with an
$\Omega_{\Lambda}=0$ Universe, and 
placed a 95\% confidence upper limit 
of $\Omega_{\Lambda}<0.7$. An $\Omega_{\Lambda}=0.7$ model predicted
about 3 times more lensed quasars than observed. This is basically
just the factor of 3 in volume between $\Omega_{\Lambda}=0$ and 
$\Omega_{\Lambda}=0.7$, shown in Fig. 1. 
They also showed that a singular isothermal
sphere (SIS) model, with the velocity dispersions of galaxies based on the
Faber-Jackson relation, gave a prediction similar to that of the
hybrid model for the number and image-separation distribution of
lensed quasars. Most of the following lensing statistics analyses
indeed used the SIS approximation. Kochanek (1996) analyzed, assuming
SIS, a somewhat 
enlarged sample obtained by adding results of severel ground-based
surveys to the Snapshot sample. He found that 5/864 quasars that are
lensed gave a    95\% confidence limit 
of $\Omega_{\Lambda}<0.66$.
  
These conclusions started changing when Chiba and Yoshii (1997,
1999) analyzed the same Snapshot sample. They  argued that, not only do
the data allow a cosmological constant, but that they actually favor
it, with a best fit $\Omega_{\Lambda}=0.7^{+0.1}_{-0.2}$ in the latter
paper. However, this result was reached by assuming a quite
``shaved'' galaxy luminosity function, with normalization $\phi^*$ cut 
by half, slope $\alpha$ changed from $-1.1$ to $+0.2$, a low-mass lens
cutoff, and $\sigma^*$, the velocity dispersion of an $L_*$ galaxy,
reduced by 20\% compared to that used by previous calculations.
At about this time, the results of two large radio surveys for lensed
quasars, JVAS and CLASS (Browne et al. 1997, 2003), also began to be
analyzed. Falco, Kochanek, \& Munoz (1998), based on a  6/2500 lensed
fraction among JVAS sources, concluded yet again that 
$\Omega_{\Lambda}$ is low, $<0.73$, at 95\% C.L. But several recent
analyses of the final combined JVAS/CLASS results, in which $\sim
12/5000$
 of the radio
sources are lensed (note that the fraction is identical to the
previous 6/2500) actually get results fully consistent with the
currently
standard cosmology: $\Omega_{\Lambda}=0.8\pm 0.1$ (Chae et al. 2002,
Chae 2003); $\Omega_{\Lambda}=0.7\pm 0.1$ (Mitchell et al. 2004).
     
\section{Where did we go wrong?}
This chain of events naturally raises the question of what went wrong
with the first studies, especially the optical ones. 
Why did lensing statistics fail to  predict the
accelerating Universe before it was discovered by other means? 
Let us examine the various observational and theoretical inputs to the
problem and attempt to locate the problem.\\
\vskip .15cm
\noindent{\bf Data?} Perhaps the results of the HST 
Snapshot Survey suffered from small number statistics, or from the
HST mirror optical aberration (even though Maoz et al. 1993 showed
this had little effect on the lensing detection efficiency).
New surveys would then show a larger lensing fraction, as predicted by
the models for the optical surveys in a $\Lambda$-dominated  Universe. However,
a second snapshot survey for lensed quasars with HST on a new sample
has found that 3/320 quasars are lensed (Morgan 2003), 
i.e., still 1\%! \\
{\bf Model Parameters?} A lensing prediction requires empirical
inputs for the properties of the lensing population. Let us examine
each of those, as they appear schematically in Eq. \ref{nsigd}.\\
{\underline{Lens number density, $n$:}} Maoz \& Rix (1993) assumed $\phi^*=1.56\times
10^{-2}$~Mpc$^{-3}$ for the density  of $L^*$ galaxies. This is nearly
identical to modern values: $\phi^*=1.59\times
10^{-2}$~Mpc$^{-3}$ (2dF; Madgwick et al. 2002) and
$\phi^*=1.49\times
10^{-2}$~Mpc$^{-3}$ (SDSS: Blanton et al. 2003). More important is the
number density of early-type galaxies, the dominant lenses. Here again    
Maoz \& Rix (1993) assumed $\phi_E^*=0.48\times
10^{-2}$~Mpc$^{-3}$, compared to: $\phi_E^*=0.45\times
10^{-2}$~Mpc$^{-3}$ (2MASS; Kochanek et al. 2001) and
$\phi_E^*=0.41\times
10^{-2}$~Mpc$^{-3}$ (SDSS: Mitchell et al. 2004). The new values
 would lower the 
predictions by only $\sim 10\%$, not by a factor of 3. The assumed logarithmic
slope of the Schechter luminosity function, $\alpha=-1.1$, was also
similar to the best current estimates for ellipticals, $\alpha\sim
-0.54$ to $-1$.\\
{\underline{Lens cross section:}} In SIS models, this parameter depends
on the  velocity dispersion of an $L^*$ early-type galaxy
as $\sigma^{*4}$. Thus,
an overestimate of $\sigma^*$ is a prime suspect for driving the lensing
predictions up, and the estimates of $\Omega_{\Lambda}$ down. The
early SIS studies assumed   $\sigma^*=225$~km~s$^{-1}$ for ellipticals     
and $206$~km~s$^{-1}$ for S0s. This compares to the modern measured values
for early types of  $209$~km~s$^{-1}$ (2MASS; Kochanek et al. 2001) and
$\sim 200$~km~s$^{-1}$ (SDSS; Sheth et al. 2003; Mitchell et
al. 2004). The latter authors actually fit a modified Schechter
function to the observed velocity dispersion distribution of SDSS
ellipticals, so a direct comparison of the ``break'' velocity
dispersion is difficult. Direct measurement of this distribution circumvents
the need to use the Faber-Jackson relation and the luminosity function
in SIS lensing statistics calculations, and is thus an important step
forward (but see Kochanek et al. 2004, for an argument that an estimate of
$\sigma^*$ from the image-separation distribution of the lens sample
itself is still superior). 
Nevertheless the peak of the cross-section-weigthed distribution 
of velocity dispersion, $\phi(\sigma)\sigma^4$, shown by Mitchell et al. (2004),
is at   $225$~km~s$^{-1}$, just the value assumed by the old studies.
A SIS calculation using $\sigma^*=200$~km~s$^{-1}$, instead of the old
values, will predict, 30-40\% fewer lensed than the old
calculations. This, on its own, cannot explain the factor 3 discrepancy.
\\
{\underline{Magnification bias $B$:}} The bias depends primarily on the source
number-magnitude relation at each redshift, or equivalently, on the 
redshift-dependent luminosity function. Maoz \& Rix (1993) used
the following parameters for the quasar luminosity function: low-luminosity
logarithmic slope $\alpha=-1.2$, high-luminosity  slope $\beta=-3.6$, 
and break absolute luminosity $M^*=-20.25$. Modern values of these
parameters (2dF, Boyle et al. 2000) are $\alpha=-1.63$, $\beta=-3.45$, 
and $M^*=-20.6$. The steepening  in $\alpha$ will tend to raise the
magnification bias. The change in $\beta$ is small. The shift in
$M^*$ to higher luminosities will lower the bias, and thus tend to
cancel the efect of a steeper $\alpha$. 
Thus, a modern bias calculation for an
optical sample would not obtain a result significantly different from
the old ones.\\
{\bf Extinction?} Extinction by dust in the lens galaxies, could, in
principle, select against lensed quasars in optical surveys, leading
to artificially low observed lensing rates. However, the magnitude of
this effect was studied by Falco et al. (1999) using the color
differences between lensed images in radio-selected and optically
selected lensed quasars. They found that dust could reduce the number
of optically selected lensed quasars by $\sim 10\%-30\%$. Again, this
effect alone would not bring the predictions for an
$\Omega_{\Lambda}=0.7$ Universe in line with the observed lensing
fraction.\\
{\bf Galaxy Evolution?} Naturally, if the lensing population were to
disappear or to become ineffective as one goes to higher redshift,
this would lower the observed number of lenses. However, Rix et al. (1994)
already showed that, if one breaks up elliptical galaxies into smaller
building blocks, as one goes to higher redshifts, and if the merging 
process occurs in a physically reasonable way, the total expected number of
lenses changes little. Since there are then more lenses along the line of
sight, but each with a lower mass, the main effect is on the image
separation distribution, with more small-separation lenses and fewer
large-separation lenses. More recently, Ofek, Rix, \& Maoz (2003) used
the observed distribution of lensing galaxy redshifts in 17 known lensed
quasar systems to limit the allowed amount of evolution in  $\sigma^*$
of early-type galaxies. They found that the lens galaxy redshifts are
consistent with no evolution out to $z=1$, independent of cosmology. At
95\% C.L., $\sigma^*(z=1)>0.63\sigma^*(z=0)$. This lack of evolution
in the elliptical population out to $z\sim 1$ is consistent also with
the results of other studies, based on number counts, colors, etc.\\
{\bf Ellipticity? Clustering?} Most lensing statistics models have
assumed circularly symmetric mass distributions for the lenses (see
Chae 2003, for an exception), and
have ignored clustering of the lenses. The impact of lens ellipticity 
and clustering 
were recently studied by Huterer, Keeton, \& Ma (2004) and Keeton \&
Zabludoff (2004), who found that their influence is small, and in the 
direction of increasing the lensing efficiency. Thus, including these
effects would actually slightly raise the predicted frequency, and
hence
would lower the deduced values of
$\Omega_{\Lambda}$.\\
{\bf Cosmology, after all?} It is evident from Fig. 1
that the volume test, which was a main
original motivation for lens surveys,
 is most powerful for very high values of $\Omega_{\Lambda}$. But
clearly, even changing $\Omega_{\Lambda}$ from the canonical 0.7 down
to, e.g., 0.6 will reduce the volume, and hence the lensing rate, by 20\%.
Such a value of $\Omega_{\Lambda}$ is still consistent with the SN-Ia
and CMB measurements. Indeed, Sullivan et al. (2003) have derived
SN-Ia Hubble diagrams, separated by morphological type of the SN host, 
based on HST imaging. For SNe-Ia in ealy-type hosts, which are least 
susceptible to extinction uncertainties, the best fit 
is $\Omega_{\Lambda}=0.5\pm 0.1$.  
    
\section{Why do the radio surveys get it ``right''?}
All of the recent lensing statistics analyses have been based on the 
JVAS/CLASS radio surveys. The rationale has been that these surveys
provide larger statistical samples, and are free of the extinction
effects that may influence optical samples. These advantages would
then lend credibility to the latest analyses, which give results
consistent with the concordance $\Omega_{\Lambda}=0.7$ cosmology. 
However, several problems with the radio surveys must be noted. First,
while the no-extinction advantage is real and important, 
it is the number of lensed systems, not the number of sources
surveyed, that determine the statistical power of the sample. There
are about a dozen
 lensed systems in the JVAS/CLASS statistical sample, 
which is similar to the 11 or 12 lenses
in the combined large optical surveys, e.g., the two HST snapshot
surveys and several ground-based  surveys (see Morgan 2003).
Second, a shortcoming of the radio surveys has been, and remains, the 
poor characterization of the source population, in terms of redshift
distribution and luminosity function. These two uncertainties lead to
uncertainties in the optical depth and in the magnification bias,
respectively. (Indeed, different assumptions about the source
population are behind the different conclusions reached from the same
observed radio lensing fraction, by Falco et al. (1998) on the one
hand, and Chae (2003) and Mitchell et al. (2004), on the other hand.)  
The poor
characterization of source redshift is illustrated by the fact that
5 of the 12 lens systems in the sample analyzed by Mitchell et
al. (2004) have unknown source redshifts. For the unlensed sources, a
single representative redshift is assumed, and the number-flux
relation is assumed to be independent of redshift. Chae (2003) has
shown that, even within the context of this limited representation of the
source population, uncertainties in the parameters have a strong
influence on the cosmological inferences. In contrast, in bright optical
surveys, the redshift of every individual source is known, and
the source number-flux relations at every redshift, and 
at fluxes much lower than those of the sample, are well characterized.
Thus, the agreement between
the calculations and the observations of lensing frequency in the 
JVAS/CLASS radio sample, assuming very similar input parameters to
those used in the old optical models, but with $\Omega_{\Lambda}=0.7$,
may be fortuitous. The agreement may result from a cancellation of
errors, where the excess of lens systems, predicted by the assumed properties
of the lensing population, is cancelled by unrealistic assumptions
made for the source population. This possibility needs to be studied
in more detail.         

\section{Conclusions}
I have argued that there was nothing particularly wrong in the 
data of the optical lens surveys of the early 90s, nor in the input
parameters of the models used to analyze them. In other words, new
optical surveys have confirmed the $\sim 1\%$ lensing fraction among
the bright quasar population, and new analyses, using updated
parameters for the source and lens populations, would reach similar 
conclusions -- namely that the observed lensing fraction is lower than
expected in a flat $\Omega_{\Lambda}=0.7$ cosmology. Since none of the 
parameters or effects in the problem can, on their own, produce a
factor of 3, or so, reduction in the observed lensing frequency, I
conclude that a conspiracy of several effects must be at work. The
most reasonable ones are that the velocity dispersion of ellipticals
are somewhat lower than assumed in past studies; that, due to extinction
by the lens galaxies, lensed systems are somewhat 
under-represented in the optical samples, compared to the
no-extinction assumption; and that, perhaps, $\Omega_{\Lambda}$ is
actually somehat less than 0.7. Each of these effects can lower the
lensing frequency by only several tens of percents, but together they can
produce the required reduction. 
I have also postulated that the agreement between the data and the models for 
radio lens surveys,
which use basically 
the same input parameters for the lensing population, may be 
fortuitous, and due to the relatively poor knowledge of the properties
of the source population.
The advantages of optical surveys in terms of characterization of the
source population mean
that a new lensing calculation of the combined optical samples, using
updated lens and source parameters, is in order. 

\begin{acknowledgments}
This work was supported by the German-Israeli Foundation for
Scientific Research and Development.
\end{acknowledgments}

\end{document}